\documentclass[article, nojss]{jss}

\usepackage{thumbpdf,lmodern}

\usepackage{framed}
\usepackage{centernot}
\usepackage{amsfonts}
\usepackage{amsmath}
\usepackage{amssymb}
\usepackage{mathtools}
\usepackage{rotating}

\newcommand{\class}[1]{`\code{#1}'}
\newcommand{\fct}[1]{\code{#1()}}
\def\ci{\perp\!\!\!\perp}

\author{Carel F.W.\ Peeters\\Amsterdam University\\Medical Centers
   \And Anders Ellern Bilgrau\\Aalborg University
   \And Wessel N. van Wieringen\\Amsterdam University\\Medical Centers}
\Plainauthor{Carel F.W. Peeters, Anders Ellern Bilgrau, Wessel N.\ van Wieringen}

\title{\pkg{rags2ridges}: A One-Stop-Shop for Graphical Modeling of High-Dimensional Precision Matrices}
\Plaintitle{rags2ridges: A One-Stop-Shop for Graphical Modeling of High-Dimensional Precision Matrices}
\Shorttitle{High-Dimensional Graphical Modeling with \pkg{rags2ridges}}

\Abstract{
  A graphical model is an undirected network representing the conditional independence properties between random variables.
  Graphical modeling has become part and parcel of systems or network approaches to multivariate data, in particular when the variable dimension exceeds the observation dimension.
  \pkg{rags2ridges} is an \proglang{R} package for graphical modeling of high-dimensional precision matrices.
  It provides a modular framework for the extraction, visualization, and analysis of Gaussian graphical models from high-dimensional data.
  Moreover, it can handle the incorporation of prior information as well as multiple heterogeneous data classes.
  As such, it provides a one-stop-shop for graphical modeling of high-dimensional precision matrices.
  The functionality of the package is illustrated with an example dataset pertaining to blood-based metabolite measurements in persons suffering from Alzheimer's Disease.
}

\Plainkeywords{graphical modeling, high-dimensional data, networks, regularization, R}

\Address{
  Carel F.W.\ Peeters (Corresponding Author)\\
  Department of Epidemiology \& Data Science\\
  Amsterdam Public Health research institute\\
  Amsterdam University Medical Centers\\
  Location VUmc\\
  1081 HV Amsterdam, the Netherlands\\
  E-mail: \email{cf.peeters@amsterdamumc.nl}\\
  URL: \url{https://cfwp.github.io/}\\
  \\
  \\
  Anders E. Bilgrau\\
  Department of Mathematical Sciences\\
  Aalborg University\\
  \emph{and}\\
  Department of Haematology\\
  Aalborg University Hospital\\
  Aalborg, Denmark\\
  URL: \url{https://bilgrau.com/}\\
  \\
  \\
  Wessel N. van Wieringen\\
  Department of Epidemiology \& Data Science\\
  Amsterdam Public Health research institute\\
  Amsterdam University Medical Centers\\
  Location VUmc\\
  \emph{and}\\
  Department of Mathematics\\
  VU University Amsterdam\\
  Amsterdam, The Netherlands\\
  URL: \url{http://www.few.vu.nl/~wvanwie/}\\
}

\begin{document}

\section{Introduction}\label{Sec:intro}

Network approaches have become ubiquitous in modern applied data science.
In an abstract sense, networks are simply a collection of nodes (or vertices) that are pairwise connected by a constellation of links (or edges).
Hence, a network is simply a graph.
What the nodes and edges signify depends on the domain of application.
In social networks for example, nodes often represent social actors such as people while the edges represent social interaction of some sort, such as friendship.
In physical networks, nodes represent physical entities such as mainframes while the edges represent physical connection such as through fiber optic cables.
In biochemical networks, nodes represent biological entities such as protein molecules, while the edges represent some form of biological interaction, such as signal transduction.
The scientific study of such networks is known as network science \citep[for an overview, see e.g.,][]{Newman10}.
Irrespective of the domain, network science provides a unifying framework for studying complex systems.

Our interest will be with networks in which the nodes represent random variables whose joint probability distribution is defined by the constellation of connections between nodes.
Such networks are known as \emph{graphical models}.
More specifically, we will consider graphs $\mathcal{G} = (\mathcal{V}, \mathcal{E})$ consisting of a finite set $\mathcal{V}$ of  $p$ vertices, corresponding to random variables $Y_1, \ldots, Y_p$
with joint probability distribution $P\sim \mathcal{N}_{p}(\mathbf{0}, \mathbf{\Sigma})$, and set of edges $\mathcal{E}$, such that for all pairs $\{Y_{j}, Y_{j'}\}$ with $j \neq j'$:
\begin{align}\nonumber
    \left(\mathbf{\Sigma}^{-1}\right)_{jj'} = (\mathbf{\Omega})_{jj'} = 0
    \Longleftrightarrow Y_{j} \ci Y_{j'}|\{Y_{j''}: j'' \neq j, j'\}
    \Longleftrightarrow (j, j') \not\in \mathcal{E}.
\end{align}
That is, a zero entry in the inverse covariance matrix (generally known as the concentration or precision matrix $\mathbf{\Omega}$) mutually implies
that the corresponding random variables are independent given the remaining variables which mutually implies that the corresponding variables are \emph{not}
connected by an undirected edge ($(j, j') \not\in \mathcal{E}$).
Hence, the undirected graph $\mathcal{G}$ is a conditional independence graph that describes the support of the (off-diagonal elements of the) precision matrix.
Such networks are generally known as \emph{Gaussian graphical models} (GGMs) and are the continuous counterpart to the Ising model from statistical physics \citep{Ising}.

The entries of the precision matrix are proportional to partial correlations.
A partial correlation is a measure of the conditional linear association between a random variable pair.
Partial correlations form the basis of network reconstruction as the conditioning weeds out spurious associations.
As such, GGMs are generally preferred to the simpler correlation networks, as available in \proglang{R} \citep{R} through the \pkg{WGCNA} package \citep{LangHor2008, WGCNA},
in which two variables are connected when their marginal correlation coefficient is nonzero.
The ability to distinguish between direct and indirect associations is especially important when the number of variables ($p$) becomes large relative to the number of observations ($n$).
In such high-dimensional data the presence of spurious associations obfuscates a mechanistic understanding of the data.

\begin{sloppypar}
The collection of high-dimensional data has become routine with the advent of high-throughput technology and data-stream mining algorithms.
A GGM is often the model of choice when the desire is to get a handle on the system of interrelations among the variables but when there is not enough
information to infer a causal structure (as indicated by a directed graph).
This has led to a flurry of packages for extracting GGMs from high-dimensional data in \proglang{R}.
The \pkg{glasso} \citep{glassopaper,glasso}, \pkg{huge} \citep{hugepaper,huge}, and \pkg{clime} \citep{climepaper,clime} packages estimate sparse precision matrices
by either penalizing the log-likelihood of the data with an $\ell_1$-penalty on the precision matrix or by exploiting the relationship between
$\ell_1$-penalized regression coefficients and precision matrix entries.
These lasso-penalized approaches are popular as $\ell_1$-penalties automatically induce sparsity.
Another approach would be to use ridge or $\ell_2$-penalization followed by post-hoc edge selection.
Such an approach underlies the \pkg{GeneNet} \citep{GeneNetpaper,GeneNet}, \pkg{GGMridge} \citep{GGMridgepaper,GGMridge}, and \pkg{beam} \citep{beampaper,beam} packages.
When the data consists of multiple classes one may consider jointly estimating multiple (sparse) precision matrices.
The \pkg{JGL} \citep{JGLpaper,JGL}, \pkg{RidgeFusion} \citep{RidgeFusion,RidgeFusionpaper}, and \pkg{pGMGM} \citep{pGMGMpaper,pGMGM} packages were created to this end.
All mentioned packages focus on network \emph{extraction} by estimating precision matrices and their support.
However, network evaluation also involves \emph{visualization} and \emph{analysis}.
Visualization pertains to the effective graphical communication of networks.
Analysis pertains to descriptions and inference on properties of retained networks.
Hence, for visualization or analysis, users of the packages above have to defer to other packages such as \pkg{igraph} \citep{igraph}.
\end{sloppypar}

\begin{figure}[ht]
\centering
\includegraphics[width=\textwidth]{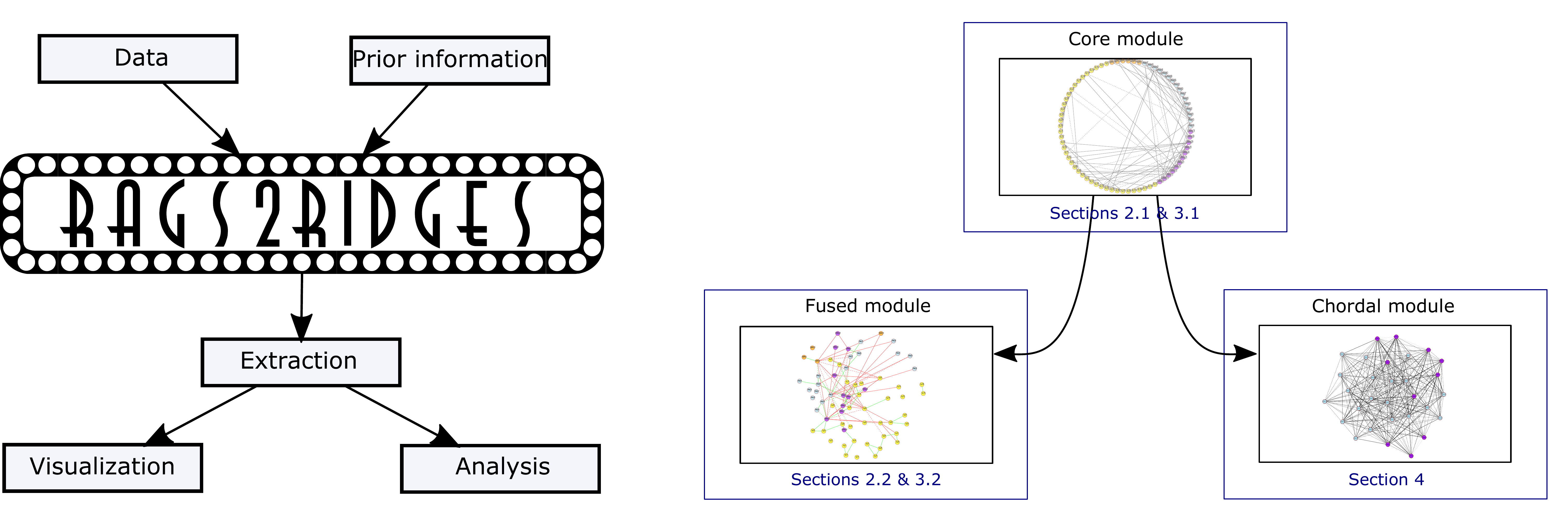}
\caption{\emph{Left-panel}: \pkg{rags2ridges} is a one-stop-shop as it (a) can combine data and prior information in inferring a GGM and (b) provides amenities for visualization and
analysis of the corresponding network.
Moreover, it (c) can do so for a single network as well as jointly for multiple networks.
\emph{Right-panel}: \pkg{rags2ridges} has a modular setup.
The core module carries workhorse functions that support and feed into extension modules.}
\label{fig:Overview}
\end{figure}

This paper introduces the \proglang{R} package \pkg{rags2ridges} \citep{rags2ridges}, available from the Comprehensive \proglang{R} Archive Network (CRAN) at
\url{https://CRAN.R-project.org/package=rags2ridges} as well as its developer page at \url{https://github.com/CFWP/rags2ridges}.
It provides functionality for the extraction as well as visualization and analysis of GGMs.
The package uses $\ell_2$-penalized maximum likelihood estimation of the precision matrix coupled with post-hoc support determination.
Such an approach has distinct advantages over a lasso-approach:
(i) It gives probabilistic control over edge selection;
(ii) It allows for the incorporation of prior information;
(iii) It can handle more extreme $p/n$ ratios \citep{WP2016};
(iv) It is more successful in retrieving the network topology when the true topology is not very sparse \citep{WP2016}.
Especially this latter point is of interest as there is accumulating evidence that many networks are dense \citep{Boyle2017}.
What sets \pkg{rags2ridges} apart from \pkg{GeneNet}, \pkg{GGMridge}, and \pkg{beam} is the usage of an algebraic $\ell_2$-penalty, resulting in an estimator that has lower risk than the
estimators in mentioned packages \citep{WP2016}.
Moreover, in contrast to all packages mentioned above, \pkg{rags2ridges} can handle the estimation of a single precision matrix, as well as the joint estimation of multiple precision matrices,
thus covering both the single and multiple group setting.
These conveniences combined make \pkg{rags2ridges} a one-stop-shop for network evaluation using GGMs (left-hand panel of Figure \ref{fig:Overview}).

While the package was originally conceived as supporting the (exploratory) extraction and analysis of biochemical networks from omics-type data,
it is certainly not restricted to that domain of application.
The package is useful whenever one wants to infer undirected graphs in which the nodes represent random variables.
As such the package has, next to transcriptomics \citep{CIBBpaper}, metabolomics \citep{deLeeuw2017}, and proteomics \citep{Hayashi2019}, also been successfully used in studies pertaining to,
for example, immune activation \citep{Blokhuis2019},  ophthalmology \citep{Elsman2020}, and primate brain shape evolution \citep{Clavel2019}.

The remainder of the paper reflects the modular setup of \pkg{rags2ridges} (right-hand panel of Figure \ref{fig:Overview}).
The core module carries the workhorse functions and revolves around the extraction, visualization, and analysis of single networks.
Section \ref{SSec:Core} handles its technicalities while Section \ref{SSec:CoreIllustrate} gives an illustration.
The core module feeds into the fused module, which revolves around the joint extraction, (differential) visualization, and analysis of multiple networks.
Section \ref{SSec:Fused} subsequently handles technicalities of the fused module while Section \ref{SSec:FusedIllustrate} gives an illustration.
The core and fused modules are the most practically useful modules.
The chordal module, as well as possible modular extensions, are discussed in Section \ref{Sec:Discuss}.

\newpage
\section{Technical details}\label{Sec:Details}
In this section we review the main technical details underlying the core (Section \ref{SSec:Core}) and fused (Section \ref{SSec:Fused}) modules.
In doing so, the main functions of the package are introduced.
Moreover, it is shown how core functions have analogues in the fused module.

\subsection{Core module}\label{SSec:Core}
\subsubsection{Penalized ML precision estimation}
Let $\boldsymbol{\mathrm{y}}_{i}$, $i=1,\ldots,n$, be a $p$-dimensional random variate drawn from $\mathcal{N}_{p}(\mathbf{0}, \mathbf{\Sigma})$.
Let $\mathbf{S}$ denote the sample covariance estimate.
The maximum likelihood (ML) estimator $\hat{\mathbf{\Omega}}$ of the precision matrix $\mathbf{\Omega} = \mathbf{\Sigma}^{-1}$ then maximizes the following log-likelihood:
\begin{align} \label{eq:form.loglik}
    \mathcal{L}(\mathbf{\Omega}; \mathbf{S})  \propto \ln|\mathbf{\Omega}| - \mbox{tr}( \mathbf{S\Omega}).
\end{align}
The maximum is achieved for $\hat{\mathbf{\Omega}}  = \mathbf{S}^{-1}$ whenever $n > p$.
It is however well-known that $\mathbf{S}$ is a poor estimate of $\mathbf{\Sigma}$ when $p\gtrapprox n$.
When $p$ approaches $n$, $\mathbf{S}$ will tend to become ill-conditioned, implying large propagation of numerical error when taking its inverse.
When $p > n$, $\mathbf{S}$ is singular, leaving $\hat{\mathbf{\Omega}}$ undefined.
We approach these problems from a penalized ML estimation perspective.

Amend the log-likelihood (\ref{eq:form.loglik}) with the $\ell_2$-penalty
\begin{align}\nonumber
    - \frac{\lambda}{2}\left\|\mathbf{\Omega} - \mathbf{T}\right\|_{2}^{2},
\end{align}
where $\mathbf{T}$ denotes a positive semi-definite symmetric target matrix and where $\lambda \in (0, \infty)$ is a strictly positive penalty parameter.
The resulting penalized log-likelihood can be maximized analytically, giving an algebraic ridge precision estimator \citep{WP2016}:
\begin{align}\label{eq:RidgeAltI}
    \hat{\mathbf{\Omega}}(\lambda) = \left\{\left[\lambda\mathbf{I}_{p}
    + \frac{1}{4}(\mathbf{S} - \lambda\mathbf{T})^{2}\right]^{1/2} +
    \frac{1}{2}(\mathbf{S} - \lambda\mathbf{T})\right\}^{-1}.
\end{align}
The estimator can be conceived as a Stein-type estimator, calibrating in the inversion the low-bias but high-variance $\mathbf{S}$ with the higher-bias but lower-variance $\mathbf{T}$.
The penalty parameter $\lambda$ then determines the amount of shrinkage.
Through $\mathbf{T}$ one can incorporate prior information without having to formally specify prior distributions.
This may range from non-informative diagonal or null-matrices to highly informative weighted adjacency matrices representing existing knowledge on the network of interest.
The function \fct{default.target} allows one to easily specify a plethora of non-informative target matrices.

The estimator in (\ref{eq:RidgeAltI}) has some appreciable properties \citep{WP2016}: (i) It is always positive-definite; (ii) It has (when $n > p$) $\mathbf{S}^{-1}$ as its right-hand limit when $\lambda \rightarrow 0$;
(iii) Its left-hand limit when $\lambda \rightarrow \infty$ is $\mathbf{T}$; (iv) It is asymptotically unbiased; and (v) It is consistent.
Moreover, under suitable penalty-choices and scalar target matrices, it uniformly dominates $\mathbf{S}$ terms of the mean squared error \citep{Wmse}.
In addition, the estimator lends itself to a natural process of iterative updating, in which the estimator from a previous round serves as $\mathbf{T}$ in a subsequent round \citep{Wjma}.
Also, it achieves lower risk than the archetypal ridge precision estimators $[(1 - \lambda^{'})\mathbf{S} + \lambda^{'}\mathbf{T}]^{-1}$ \citep[see, e.g.,][]{LWest,GeneNetpaper,beampaper} and
$[\mathbf{S} + \lambda^{''}\mathbf{I}_{p}]^{-1}$ \citep[see, e.g.,][]{GGMridgepaper}, both of which are also available in \pkg{rags2ridges}.
As such, the estimator (\ref{eq:RidgeAltI}) provides an adequate basis for graphical modeling.
It is available through the workhorse function \fct{ridgeP} when using the default argument \code{type = "Alt"}.

\subsubsection{Choosing the penalty parameter}
An important issue is choosing a value for the penalty parameter.
One wants to choose $\lambda$ such that the precision estimate is well-conditioned while not unnecessarily suppressing relevant data signal.
As the $\ell_2$-penalty does not automatically induce sparsity in the estimate, it is natural to choose a method that seeks loss efficiency rather than model selection consistency.
\pkg{rags2ridges} employs cross-validation (CV).
The $K$-fold CV score for the estimate $\hat{\mathbf{\Omega}}(\lambda)$ based on the fixed penalty $\lambda$ can be given as:
\begin{align}\nonumber
    \varphi_{K}(\lambda) = \sum_{k=1}^{K} n_k \left\{
    -\ln|\hat{\mathbf{\Omega}}_{-k}(\lambda)| +
    \mbox{tr}[\mathbf{S}_k\hat{\mathbf{\Omega}}_{-k}(\lambda)] \right\},
\end{align}
where $n_k$ is the size of subset $k$, for $k = 1, \ldots, K$ disjoint subsets.
Further, $\mathbf{S}_k$ denotes the sample covariance matrix estimate based on subset $k$, while
$\hat{\mathbf{\Omega}}_{-k}(\lambda)$ denotes the estimated regularized precision matrix on all samples not in $k$.
The optimal penalty value $\lambda^*$ is then chosen such that:
\begin{align}\nonumber
    \lambda^* = \arg\min_{\lambda}{\varphi}_{K}(\lambda),
\end{align}
which relates to the maximization of predictive accuracy.
The \fct{optPenalty} family of functions gives access to penalty value determination by evaluation of the $K$CV negative log-likelihood score.
It contains brute-force, approximation, and automated-search options.
The automated-search option applies a root-finding algorithm to automatically search for the optimal value along the domain of $\lambda$ and is our usual option of choice.

After obtaining $\hat{\mathbf{\Omega}}(\lambda^{*})$, we might like to assess its conditioning.
We use the spectral condition number, i.e., the ratio of the maximum to minimum eigenvalue, for this assessment.
A matrix with a high condition number is said to be ill-conditioned.
Numerically, ill-conditioning will mean that small changes in the penalty parameter lead to dramatic changes in the condition number.
When plotting the condition number against the (domain of the) penalty parameter, there is a point of relative stabilization (when working in the $p > n$ situation)
that can be characterized by a leveling-off of the acceleration along the curve \citep{CNplot}.
Such condition number plots are the output of the \fct{CNplot} function.
The condition number plot thus tracks the domain of the penalty parameter for which the regularized precision matrix is ill-conditioned and
visualizes whether $\hat{\mathbf{\Omega}}(\lambda^{*})$ finds itself in the stable domain of well-conditionedness.

\subsubsection{Support determination}
If we have a well-conditioned estimate of the precision matrix we want to extract a network.
In graphical modeling this implies determining the support of the estimated (regularized) precision matrix.
Support determination is best performed on a scaled version of the precision matrix: the partial correlation matrix $\hat{\mathbf{P}}(\lambda^{*})$.
The support of the partial correlation matrix equals the support of the corresponding precision matrix, but the former simplifies evaluation by having all entries on the same scale.
The \fct{sparsify} function determines the support of a partial correlation/precision matrix by post-hoc thresholding and sparsifies it accordingly.

The function offers multiple options for thresholding.
The most simple is available through the argument \code{threshold = "top"}.
It allows one to retain the $x$ strongest edges in terms of their absolute partial correlation values by additionaly specifying \code{top = x}.
Another simple option is accessed through \code{threshold = "absValue"}, which retains all edges whose corresponding absolute partial correlation
$|[\hat{\mathbf{P}}(\lambda^{*})]_{jj'}| \geq$ \code{absValueCut}.
The \fct{sparsify} function also offers probabilistic control over edge selection through a multiple testing procedure.
Specifying the argument \code{threshold = "localFDR"} leads to support determination by a local false discovery rate (lFDR) procedure proposed by \cite{GeneNetpaper}.
It assumes that the nonredundant partial correlation coefficients follow a mixture distribution:
\begin{align}\label{eq:MIx}
    f\left\{ [\hat{\mathbf{P}}(\lambda^{*})]_{jj'} \right\} =
    \eta_{0} f_{0} \left\{ [\hat{\mathbf{P}}(\lambda^{*})]_{jj'} ; \kappa \right\} +
    (1 - \eta_{0}) f_{\mathcal{E}} \left\{ [\hat{\mathbf{P}}(\lambda^{*})]_{jj'} \right\},
\end{align}
with mixture weight $\eta_{0} \in [0,1]$.
In (\ref{eq:MIx}), $f_{0}\{\cdot\}$ denotes the distribution of a null-edge while, which is a scaled beta-density with $\kappa$ degrees of freedom.
The distribution of a present edge is given by $f_{\mathcal{E}}\{\cdot\}$.
In the $p > n$ situation one has to estimate $\kappa$, $\eta_{0}$ and $f_{\mathcal{E}}\{\cdot\}$, the details of which can be found in \cite{Efron2010} and \cite{GeneNetpaper}.
With these estimates the lFDR can be calculated, representing the empirical posterior probability that $Y_{j}$ and $Y_{j'}$ are \emph{not} connected given their observed partial correlation:
\begin{align}\nonumber
    P\left((j, j') \not\in \mathcal{E} | [\hat{\mathbf{P}}(\lambda^{*})]_{jj'} \right) =
    \frac{\hat{\eta}_{0} f_{0} \left\{ [\hat{\mathbf{P}}(\lambda^{*})]_{jj'} ; \hat{\kappa} \right\}}
    {\hat{\eta}_{0} f_{0} \left\{ [\hat{\mathbf{P}}(\lambda^{*})]_{jj'} ; \hat{\kappa} \right\} +
    (1 - \hat{\eta}_{0}) \hat{f}_{\mathcal{E}} \left\{ [\hat{\mathbf{P}}(\lambda^{*})]_{jj'} \right\}}.
\end{align}
The procedure then will retain those edges whose probability of being present equals or exceeds the probability specified in the \code{FDRcut} argument, i.e.,
it will retain those edges for which $1 - P\left((j, j') \not\in \mathcal{E} | [\hat{\mathbf{P}}(\lambda^{*})]_{jj'} \right) \geq$ \code{FDRcut}.

\subsubsection{Visualization}
The \fct{Ugraph} function is the main function for the network visualization of the sparsified precision matrix $\hat{\mathbf{\Omega}}(\lambda^{*})_{0}$
or the sparsified partial correlation matrix $\hat{\mathbf{P}}(\lambda^{*})_{0}$.
Visualization is very important for a first understanding of a network.
The very same network can be visualized in uncountably many ways, and some ways might opaque rather than enlighten its structure and properties.
Hence, the \fct{Ugraph} function tries to offer flexibility as well as easy access to variety in graph drawing.
Through \fct{Ugraph} one can control, among others, node size, node color, edge color, edge thickness, and node-pruning (trimming nodes that are unconnected).
Control over these elements is essential when the network has many nodes and/or many connections.
Moreover, \fct{Ugraph} allows for the differentiation between positive and negative edge weights.
Also, the function gives a plethora of options for (force-directed) layout algorithms.
Most layout functions supported by \pkg{igraph} are supported (the function is partly a wrapper around certain \pkg{igraph} functions).
These layouts can be invoked by a character that mimics a call to an \pkg{igraph} layout function through the \code{lay} argument.
When using \code{lay = NULL} one can specify the placement of nodes with the \code{coords} argument.
The row dimension of this matrix should equal the number of (pruned) vertices.
The column dimension then should equal 2 (for 2D layouts) or 3 (for 3D layouts).
The \code{coords} argument can also be viewed as a convenience argument as it enables one, e.g., to layout a graph according to the coordinates of a previous call to \fct{Ugraph}.
This enables the visual comparison of multiple graphs.

\subsubsection{Analysis}
Following visualization we might desire analyzing the retained structure.
The simplest analyzes describe network properties at the node-level.
Dominant in this respect are \emph{centrality measures}, which aim to identify the most important nodes (hubs) within a network.
There are many centrality measures that capture different aspects of node importance.
Two well-known and oft studied centrality measures are degree centrality and betweenness centrality.
Degree centrality simply indicates the number of connections in which a node takes part.
It is indicative of the nodes that are central or influential in terms of the number of connections: more connections could imply deeper regulatory influence.
Betweenness centrality measures centrality in terms of information flow.
Under the assumption that information is passed over short(est) paths a node becomes central when the number of short(est) paths that pass through it is high.
The \fct{GGMnetworkStats} function calculates, for a given sparsified precision or partial correlation matrix, the degree and betweenness centralities as well as various other measures of centrality.
It also calculates the number of positive and the number of negative edges for each node.
In addition, for each variate the mutual information (with all other variates), the variance, and the partial variance is represented.
See, e.g., \cite{Newman10} for more information on centrality measures and other node-level statistics.

We might also analyze network properties from the perspective of paths.
A path is a sequence of edges that connect a sequence of distinct nodes.
One might then be interested, for example, in the collection of strongest paths between two endnodes of interest.
The \fct{GGMpathStats} function uses a result by \cite{JW05} stating that the observed covariance between any pair of nodes can be decomposed as a sum of path contributions for all paths connecting these
nodes in the conditional independence graph.
The function returns all paths, their respective lengths, and their respective contributions to the marginal covariance between the endnodes of interest.
It allows one to identify, for example, the strongest paths between such nodes, and if these are mediating or moderating paths.

Another analysis perspective is found in groups of nodes.
The nodes in a network often cluster in groups: collections of nodes that are more deeply connected to each other than to nodes outside their direct topological environment.
These groups are often called \emph{communities}.
There is interest in the detection of these communities as they are thought to represent functional modules.
Community detection, loosely speaking, then refers to ``the search for naturally occurring groups in a network" \citep[][p.\ 371]{Newman10}.
The \fct{Communities} function performs community detection.
It uses an edge-betweenness-based method of community detection, commonly known as the Girvan-Newman algorithm \citep{GN04}.
The community structure in the graph can also be visualized when using the argument \code{graph = TRUE}.
The arguments for this visualization are analogous to those in the \fct{Ugraph} function.

\subsection{Fused module}\label{SSec:Fused}
Often, data consists of observations that are subject to class membership.
Class membership may have different connotations.
It may refer to certain sub-classes within a single data set such as disease subtypes.
It may also designate different data sets or studies.
Likewise, ``the class indicator may also refer to a conjunction of both subclass and study membership to form a two-way design of factors of interest
(e.g., breast cancer subtypes present in a batch of study-specific data sets)'' \citep{fused}.
In such settings one may desire the joint estimation of multiple regularized precision matrices from (aggregated) high-dimensional data consisting of distinct classes.
The fused module in \pkg{rags2ridges} contains functions for doing so.

Suppose $\boldsymbol{\mathrm{y}}_{ig}$ is a realization of a $p$-dimensional random variate for $i = 1, \ldots, n_g$ independent observations nested within $g = 1, \ldots, G$ non-overlapping classes,
drawn from $Y_{g} \sim \mathcal{N}_p(\boldsymbol{0}, \mathbf{\Sigma}_g)$.
Our desire is to obtain an estimate of the precision matrix $\mathbf{\Omega}_g$ for each class $g$.
Let $\{\mathbf{\Omega}_g\}$ and $\{\mathbf{S}_g\}$ respectively denote the sets $\{\mathbf{\Omega}_g\}_{g=1}^{G}$ and $\{\mathbf{S}_g\}_{g=1}^{G}$, and consider the log-likelihood for
the joint $n_{\bullet} = \sum_{g = 1}^G n_g$ data observations:
\begin{align}\label{eq:joint.form.loglik}
  \mathcal{L}\left(\{\mathbf{\Omega}_g\}; \{\mathbf{S}_g\}\right)
    \propto \sum_g n_g
      \Big\{ \ln|\mathbf{\Omega}_g| - \mbox{tr}(\mathbf{S}_g\mathbf{\Omega}_g) \Big\}.
\end{align}
We again consider a high-dimensional setting in which $p > n_{\bullet}$.
In such a situation the class-specific ML estimates $\mathbf{S}_g$ are singular and their inverses do not exist.
Moreover, if they could be obtained then the precision estimates would be treated as if they exist in independent class-specific vacuums.
The pooled covariance matrix $n_{\bullet}^{-1} \sum_{g = 1}^G n_g \mathbf{S}_g$, often used as an estimate of the common covariance matrix across classes, will also
be singular with an undefined inverse.
In addition, if this inverse would exist, it ignores class-specific information.
These problems can be approached by generalizing the $\ell_2$-penalized ML framework from Section \ref{SSec:Core} to \emph{targeted fused ridge estimation}.

Amend, to this end, the log-likelihood (\ref{eq:joint.form.loglik}) with the following fused $\ell_2$-penalty \citep{fused}:
\begin{align}\label{eq:FR}
     - \sum_g \frac{\lambda_{gg}}{2} \left\|{\mathbf{\Omega}_g {-} \mathbf{T}_g}\right\|_{2}^{2}
      +  \sum_{\mathclap{g_1, g_2}} \frac{\lambda_{g_1 g_2}}{4}
        \left\|(\mathbf{\Omega}_{g_1} {-} \mathbf{T}_{g_1}) {-}
                          (\mathbf{\Omega}_{g_2} {-} \mathbf{T}_{g_2})\right\|_{2}^{2},
\end{align}
where the $\mathbf{T}_g$ denote class-specific positive semi-definite symmetric target matrices,
the $\lambda_{gg} \in (0,\infty)$ denote class-specific ridge penalty parameters,
and where the $\lambda_{g_1 g_2} = \lambda_{g_2 g_1} \in [0,\infty)$ indicate pair-specific \emph{fusion} penalty parameters.
All penalties can be collected in the penalty matrix $\mathbf{\Lambda}$, carrying the $\lambda_{gg}$ along the diagonal and the $\lambda_{g_1 g_2}$ off-diagonal.
The penalty $\lambda_{gg}$ controls the rate of shrinkage of $\mathbf{\Omega}_g$ towards $\mathbf{T}_g$ while $\lambda_{g_1 g_2}$
determines the retainment of entry-wise similarities between $(\mathbf{\Omega}_{g_1} {-} \mathbf{T}_{g_1})$ and $(\mathbf{\Omega}_{g_2} {-} \mathbf{T}_{g_2})$.
The addition of the fused $\ell_2$-penalty ``allows for the joint estimation of multiple precision matrices from high-dimensional data classes that chiefly
share the same structure but that may differentiate in locations of interest" \citep[][p.\ 5]{fused}.
It also improves sensitivity and specificity of topological discoveries by borrowing power across classes \citep{fused}.
The formulation of the penalty in (\ref{eq:FR}) is very general and can be specialized to many special cases, including the single-network setting of Section \ref{SSec:Core}.

The targeted fused penalized log-likelihood has a closed-form maximizing argument for any given class $g$ \citep{fused}:
\begin{align}\label{eq:updateFused}
  \hat{\mathbf{\Omega}}_{g}\bigl(\mathbf{\Lambda}, \{\mathbf{\Omega}_{g'}\}_{g' {\neq} g} \bigr)
  =
  \left\{
    \left[
      \bar{\lambda}_{g} \mathbf{I}_p
      + \frac{1}{4}\big(\bar{\mathbf{S}}_{g} - \bar{\lambda}_{g}{\mathbf{T}}_{g}\big)^2
    \right]^{1/2}
    + \frac{1}{2}\big(\bar{\mathbf{S}}_{g} - \bar{\lambda}_{g} {\mathbf{T}}_{g}\big)
  \right\}^{-1},
\end{align}
which can be viewed as a generalization of (\ref{eq:RidgeAltI}), with
\begin{align}\nonumber
  \bar{\mathbf{S}}_{g}
  = \mathbf{S}_{g} - \sum_{g' \neq g}\frac{\lambda_{g' g}}{n_{g}} ({\mathbf{\Omega}}_{g'} \!- {\mathbf{T}}_{g'}),
  \label{eq:barupdate}
  \quad\text{and}\quad
  \bar{\lambda}_{g} = \frac{\sum_{g'} \lambda_{g' g}}{n_{g}}.
\end{align}
The estimator (\ref{eq:updateFused}) has appreciable properties analogous to the properties stated for (\ref{eq:RidgeAltI}).
It forms the iterative core of finding the general maximizing argument $\{\hat{\mathbf{\Omega}}_g\}$ by block coordinate ascent \citep{fused}.

The block coordinate ascent procedure can itself be subjected to cross-validation to select values for the entries in $\mathbf{\Lambda}$.
In fact, the full functionality of the core module can be extended to deal with the joint estimation and evaluation of multiple (sparsified) precision matrices.
The functions of this fused module, suffixed with \code{.fused}, mirror the functions of the core module.
Table \ref{Tab:Overview} gives an overview of core functions and their fused analogues as well as their usage in the network evaluation cycle: extraction, visualization, analysis.
Most functions have \proglang{C++} at their core \citep{rcpp}, ensuring (relative) computational swiftness \citep[see, e.g., Section 5.6 of][]{fused}.

\begin{table}[b!]
\centering
\begin{tabular}{lll}
\hline
Core function           & Fused analogue                    & Supports \\ \hline
\fct{ridgeP}            & \fct{ridgeP.fused}                & Extraction       \\
\fct{default.target}    & \fct{default.target.fused}        & Extraction       \\
\fct{optPenalty}        & \fct{optPenalty.fused}            & Extraction       \\
\fct{sparsify}          & \fct{sparsify.fused}              & Extraction       \\
\fct{Ugraph}            & \fct{DiffGraph}                   & Visualization    \\
\fct{GGMnetworkStats}   & \fct{GGMnetworkStats.fused}       & Analysis         \\
\fct{GGMpathStats}      & \fct{GGMpathStats.fused}          & Analysis         \\
\fct{Communities}       & -                                 & Analysis         \\ \hline
\end{tabular}
\caption{Overview of the main functions in the \pkg{rags2ridges} package.}
\label{Tab:Overview}
\end{table}


\section{Illustrations} \label{Sec:illustrations}
Section \ref{SSec:CoreIllustrate} illustrates the extraction, visualization, and analysis of \emph{single} networks through the core module.
Section \ref{SSec:FusedIllustrate} illustrates the joint extraction, visualization, and analysis of \emph{multiple} networks through the fused module.
Throughout, we will be using metabolomics data on patients with Alzheimer's Disease (AD).
AD is a chronic neurodegenerative disease and the main cause of dementia.
Metabolomics refers to the collective quantification of metabolites: small-molecules that result from metabolic processes.
Metabolomics on pheripheral fluids is of interest in terms of non-invasive measurement of disease-specific biochemical changes.

The data, internally packaged for educational purposes, are a doubly anonymized subset of the data reported in \citet{deLeeuw2017}.
The data can be loaded with a simple call.
\begin{CodeChunk}
\begin{CodeInput}
R> data(ADdata)
\end{CodeInput}
\end{CodeChunk}
The \class{ADdata} consists of 3 objects related to blood-based metabolomics data on various AD subtypes.
\class{ADmetabolites} is a matrix containing metabolic expressions of $p = 230$ metabolites (rows) on $n = 127$ samples (columns).
The \class{sampleInfo} object is a \class{data.frame} containing clinical information on the samples.
This information pertains to diagnosis: AD Class 1 (AD patients without a known genetic predisposition for AD) or AD Class 2 (AD patients with a known genetic predisposition for AD).
The \class{variableInfo} object is also a \class{data.frame}, containing information on the metabolic features.
This information pertains to the compound family a metabolite belongs to: Amines, organic acids, lipids, or oxidative stress compounds.
These objects are easily probed by calling \code{head(ADmetabolites)}, \code{head(sampleInfo)}, and \code{table(variableInfo)}, providing basic insights into metabolite expression, and clinical and feature information.

\subsection{Core module} \label{SSec:CoreIllustrate}

\subsubsection{Extraction}
We first need an estimate of the precision matrix.
In this section we will be interested in data for AD Class 2 (i.e, AD patients with a known genetic predisposition for AD).
The precision matrix estimate will have to be a regularized estimate as the number of features ($p = 230$) is larger than the number of observations ($n = 87$ for AD Class 2).
We will use the ridge estimate of the precision matrix given in (\ref{eq:RidgeAltI}).

We start by scaling the data, as the features have different scales and as the variability of the features may differ substantially.
\begin{CodeChunk}
\begin{CodeInput}
R> ADclass2 <- ADmetabolites[,sampleInfo$ApoEClass == "Class 2"]
R> ADclass2 <- scale(t(ADclass2))
\end{CodeInput}
\end{CodeChunk}
One could simply call \fct{ridgeP} if one would desire a ridge estimate of the precision matrix for a \emph{given} penalty value (say, e.g., $\lambda = .3$).
\begin{CodeChunk}
\begin{CodeInput}
R> P <- ridgeP(covML(ADclass2), lambda = .3)
\end{CodeInput}
\end{CodeChunk}
The (scaled) data can also be used in the \fct{optPenalty.kCVauto} function which uses \fct{ridgeP} internally.
This function determines the optimal value of the penalty parameter by application of the Brent algorithm \citep{Brent71} to the $K$CV negative log-likelihood score.
The search for the optimal value is automatic in the sense that in order to invoke the root-finding abilities of the Brent method,
only a minimum value (\code{lambdaMin}) and a maximum value (\code{lambdaMax}) for the penalty parameter need to be specified (from which a starting penalty value is automatically generated).
As we have little \emph{a priori} knowledge, we will keep the target matrix non-informative: An identity matrix (one of the default targets in the \fct{default.target} function).
We choose $K = 10$, such that the penalty value at which the 10-fold cross-validated negative log-likelihood score is minimized is deemed optimal.
\begin{CodeChunk}
\begin{CodeInput}
R> set.seed(1234)
R> OPT <- optPenalty.kCVauto(
+    ADclass2,
+    fold      = 10,
+    lambdaMin = 1e-07,
+    lambdaMax = 20,
+    target    = default.target(covML(ADclass2), type = "DUPV")
+  )
\end{CodeInput}
\end{CodeChunk}
The output is an object of class \class{list}, with \code{$optLambda} containing the optimal penalty value and \code{$optPrec} containing the precision
matrix under the optimal value of the penalty parameter.
It is often informative to have a look at the optimal penalty value.
\begin{CodeChunk}
\begin{CodeInput}
R> OPT$optLambda
\end{CodeInput}
\begin{CodeOutput}
0.1561957
\end{CodeOutput}
\end{CodeChunk}

\begin{figure}[t!]
\centering
\includegraphics[width=\textwidth]{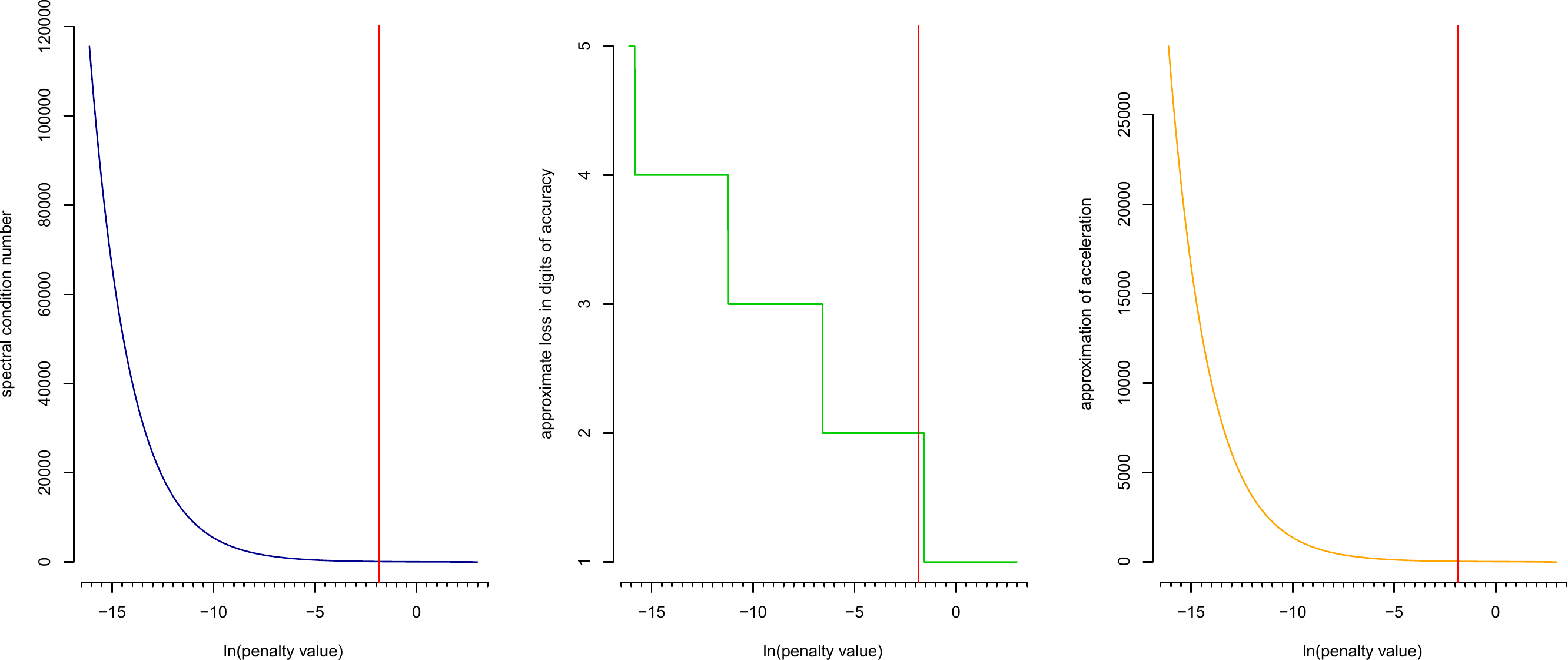}
\caption{\label{fig:CNplot} A condition number plot with interpretational aids.}
\end{figure}

Using the \fct{CNplot} function, we can assess the conditioning of $\hat{\mathbf{\Omega}}(.1561957)$.
It outputs a graph of the spectral condition number over the domain of the penalty parameter.
\begin{CodeChunk}
\begin{CodeInput}
R> CNplot(covML(ADclass2),
+         lambdaMin = 1e-07,
+         lambdaMax = 20,
+         step      = 5000,
+         target    = default.target(covML(ADclass2), type = "DUPV"),
+         Iaids     = TRUE,
+         vertical  = TRUE,
+         value     = OPT$optLambda)
\end{CodeInput}
\end{CodeChunk}
When \code{Iaids = TRUE}, the basic condition number plot is amended/enhanced with two additional plots (over the same domain of the penalty parameter as the basic plot):
(i) The approximate loss in digits of accuracy (for the operation of inversion) and (ii) an approximation to the second-order derivative of the curvature in the basic plot.
These interpretational aids can enhance interpretation of the basic condition number plot \citep{CNplot}.
When \code{vertical = TRUE} a vertical line is added at the constant given in the argument \code{value}.
This option can be used to assess if the previously obtained optimal penalty value has led to a precision estimate that is well-conditioned.
The graph is given in Figure \ref{fig:CNplot}.
The far-left panel gives the basic spectral condition number plot.
The red vertical line indicates the value for the penalty that was deemed optimal.
This optimal penalty value lies in the domain of well-conditionedness: small perturbations in this value do not affect the spectral condition number greatly.
The middle and far-right panels depict the interpretational aids.
From the middle plot, for example, we understand that the estimated loss in digits of accuracy in operations based on the optimal precision matrix is $2$.

Now that we have a well-conditioned estimate of the precision matrix we want to determine its support.
We will use the lFDR option in the \fct{sparsify} function for doing so.
Elements are retained if their posterior probability of being present (equalling $1 -$ lFDR) $\geq$ \code{FDRcut}.
As metabolic networks are very dense, we set \code{FDRcut} quite high: $.999$.
\begin{CodeChunk}
\begin{CodeInput}
R> P0 <- sparsify(OPT$optPrec,
+                 threshold = "localFDR",
+                 FDRcut    = .999,
+                 verbose   = FALSE)
\end{CodeInput}
\begin{CodeOutput}
- Retained elements:  310
- Corresponding to 1.18 % of possible edges
\end{CodeOutput}
\end{CodeChunk}
The function returns an object of class \class{list}:
\code{P0$sparsePrecision} contains the sparsified precision matrix while \code{P0$sparseParCor} contains its scaled version, the sparsified partial correlation matrix.

\begin{figure}[t!]
\centering
\includegraphics[width=.55\textwidth]{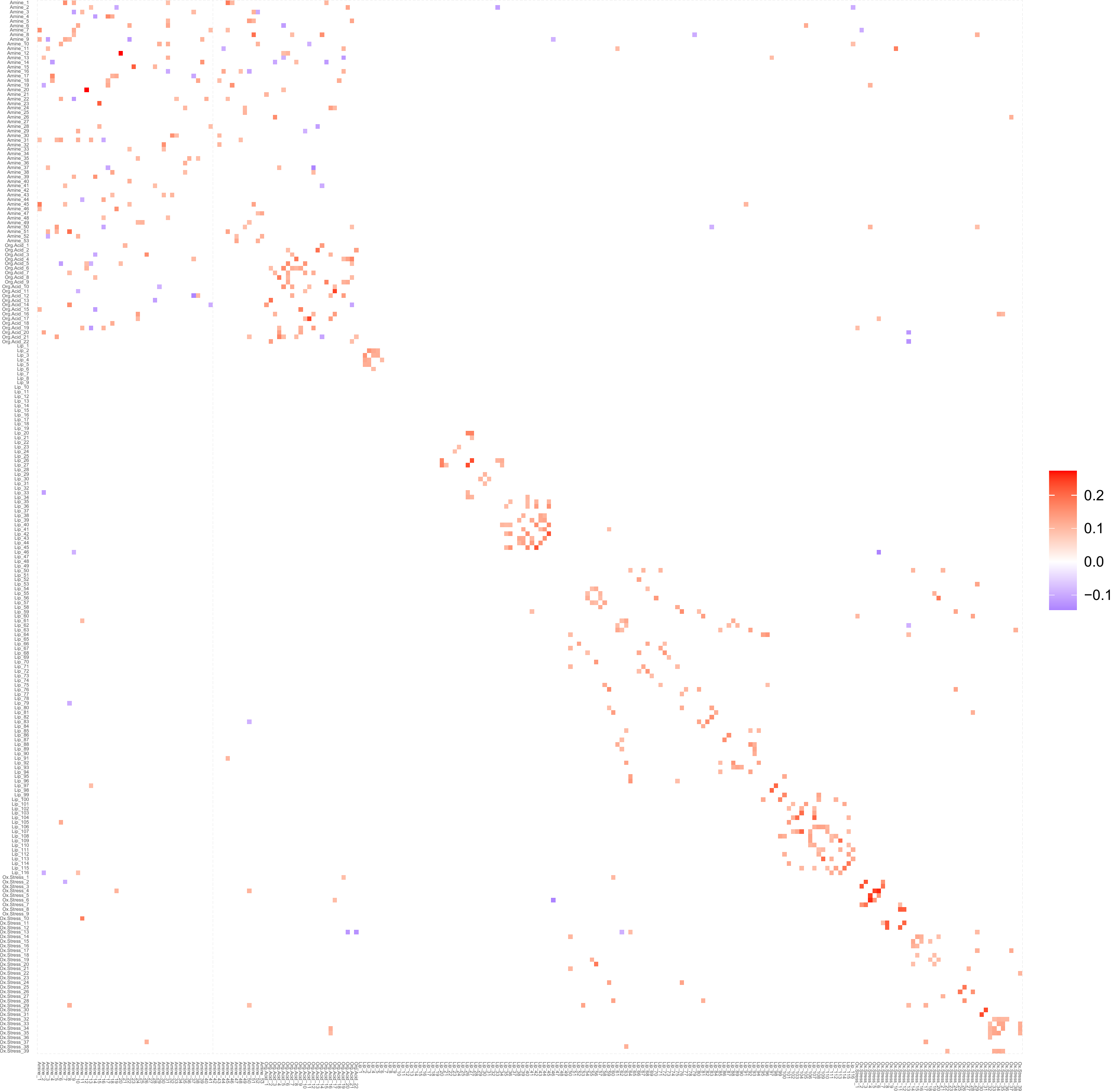}
\caption{\label{fig:Heat} Heatmap of the sparsified partial correlation matrix.
The legend represents the color scale for entries in the negative (blue) and positive (red) range.}
\end{figure}

A first visualization of the sparsified matrices is easily obtained through usage of \fct{edgeHeat} function.
\begin{CodeChunk}
\begin{CodeInput}
R> edgeHeat(P0$sparseParCor, diag = FALSE, textsize = 3)
\end{CodeInput}
\end{CodeChunk}
The resulting heatmap, given in Figure \ref{fig:Heat}, is essentially a visualization of the extracted network, as it represents the weighted adjacency matrix of the conditional independence graph
at hand.
However, more insightful network visualization is possible, a topic to which we turn next.

\subsubsection{Visualization}
\begin{figure}[h!]
\centering
\includegraphics[width=\textwidth]{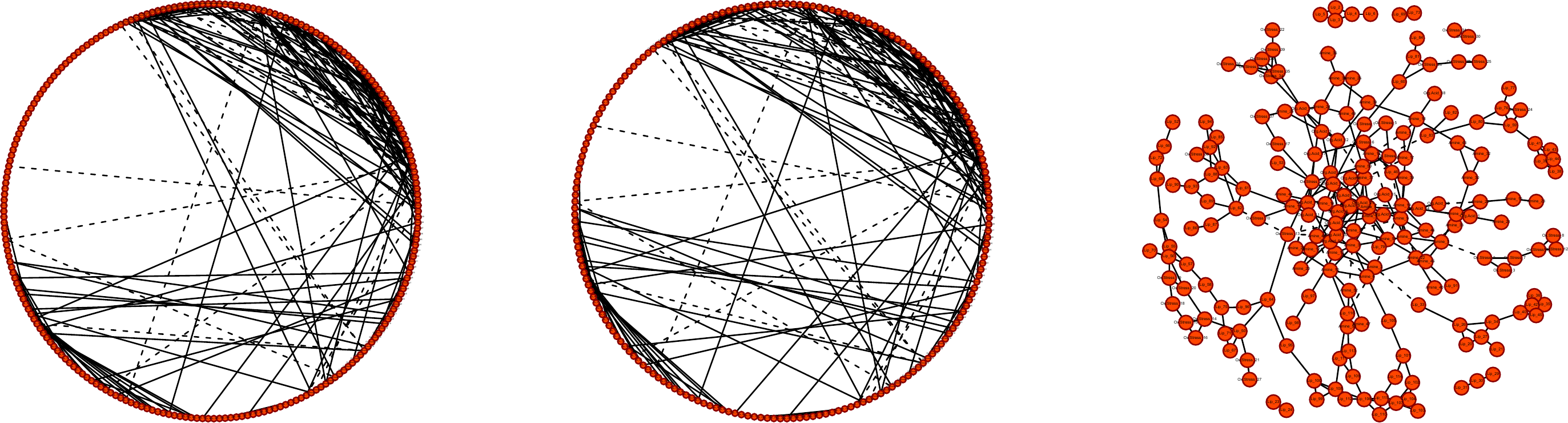}
\caption{\label{fig:SimpleViz} Visual results from various simple calls to \fct{Ugraph}.}
\end{figure}

Visualization is very important for a first understanding of the graph.
Visualization should always be geared towards maximizing information and clarity with a minimum of (visual) clutter.
The code below gives increasingly elaborate calls to \fct{Ugraph}.
\begin{CodeChunk}
\begin{CodeInput}
R> par(mfrow=c(1,3))
R> Ugraph(P0$sparseParCor,
+         type  = "fancy",
+         Vsize = 3,
+         Vcex  = .1)
R> Ugraph(P0$sparseParCor,
+         type  = "fancy",
+         Vsize = 3,
+         Vcex  = .1,
+         prune = TRUE)
R> set.seed(1567)
R> Coords <- Ugraph(P0$sparseParCor,
+                   type  = "fancy",
+                   lay   = "layout_with_fr",
+                   prune = TRUE,
+                   Vsize = 7,
+                   Vcex  = .3)
\end{CodeInput}
\end{CodeChunk}
The first call only specifies the type of graph (\code{type}), the size of the vertices (\code{Vsize}), and the size of the vertex labels (\code{Vcex}).
When \code{type = "fancy"} a graph is given in which dashed lines indicate negative edges while solid lines indicate positive edges.
The result (left-hand panel of Figure \ref{fig:SimpleViz}) is a graph/network under a default layout that gives a circular placement of the vertices.
There are quite some features, such that the network structure is not directly clear.
One could try to clarify the structure by pruning the graph.
That is, by removing unconnected nodes.
This is easily achieved by adding the \code{prune = TRUE} argument, given in the second call.
As there are few unconnected nodes, the pruning does little to alleviate the clutter (middle-panel of Figure \ref{fig:SimpleViz}).
In this situation it might be better to use a layout different from the circular placement of vertices.
One option is the (force-directed) Fruchterman-Reingold algorithm \citep{FR91} which tries to position the nodes such that all edges
are approximately of equal length and tries to minimize the number of crossing edges.
Such a layout can be obtained by specifying \code{lay = "layout_with_fr"}, given in de third call.
The structure of the network now becomes more clear as the new layout seems to suggest that there are core and pheriphery substructures (rights-hand panel of Figure \ref{fig:SimpleViz}).

\begin{figure}[h!]
\centering
\includegraphics[width=\textwidth]{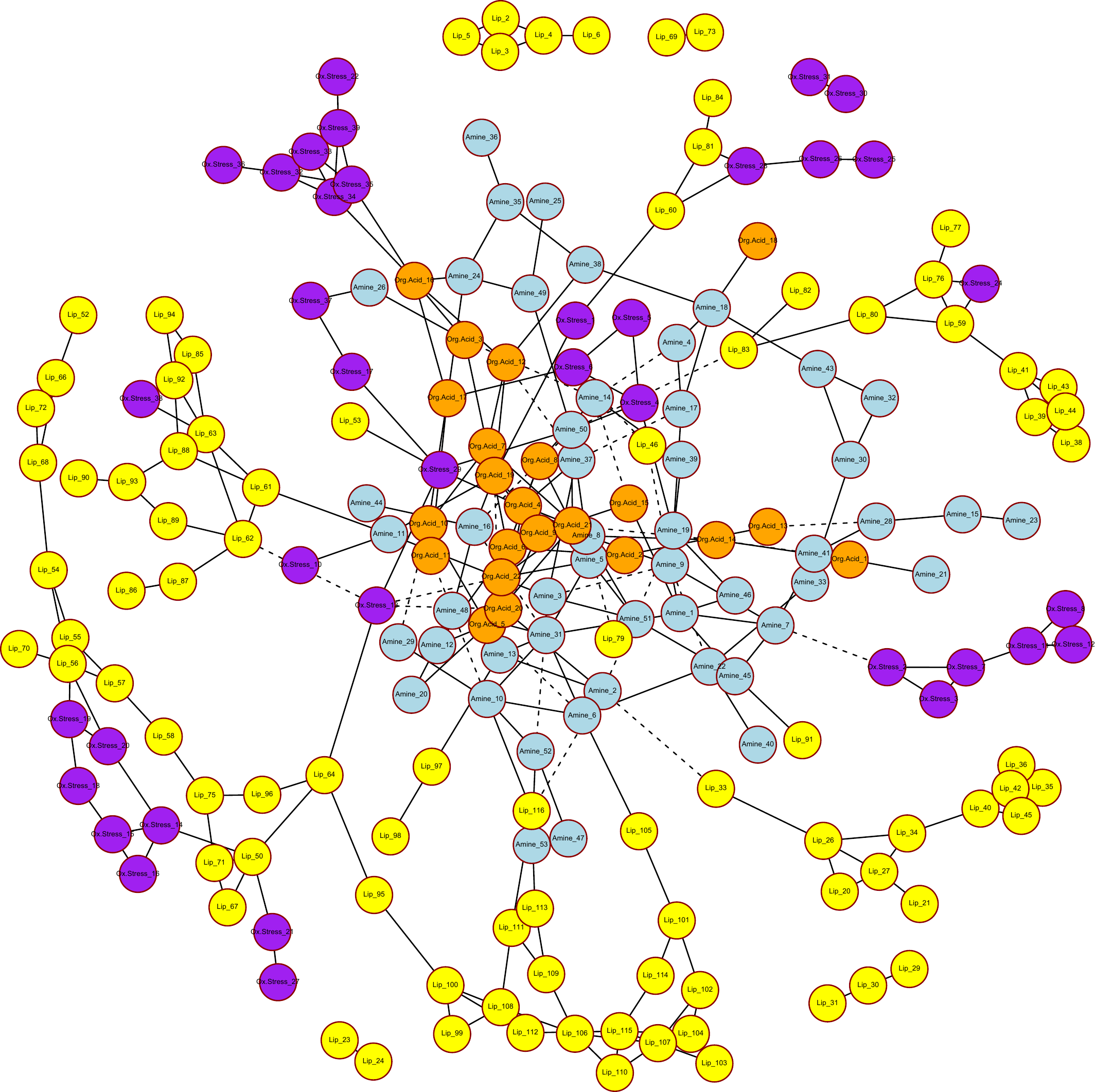}
\caption{\label{fig:NetW} Metabolite network for AD patients with a known genetic predisposition for AD.
Dashed edges indicate negative edge weight.
Solid edges indicate positive edge weight.
Node colorings are according to compound family: blue for amines, orange for organic acids, yellow for lipids, and purple for oxidative stress compounds.}
\end{figure}

Even more information can be conveyed if we would include information on compound families in the node-colorings.
That is, we could color each node according to the compound family it belongs to.
The \fct{Ugraph} function also supports visual comparison of networks as it outputs the layout-coordinates of the network that is visualized.
Hence, for the next call to \fct{Ugraph} the retained coordinates of the previous call are used for node-placement.
\begin{CodeChunk}
\begin{CodeInput}
R> PcorP  <- pruneMatrix(P0$sparseParCor)
R> Colors <- rownames(PcorP)
R> Colors[grep("Amine", rownames(PcorP))]     <- "lightblue"
R> Colors[grep("Org.Acid", rownames(PcorP))]  <- "orange"
R> Colors[grep("Lip", rownames(PcorP))]       <- "yellow"
R> Colors[grep("Ox.Stress", rownames(PcorP))] <- "purple"

R> par(mfrow=c(1,1))
R> Ugraph(PcorP,
+         type   = "fancy",
+         lay    = NULL,
+         coords = Coords,
+         Vcolor = Colors,
+         Vsize  = 7,
+         Vcex   = .3)
\end{CodeInput}
\end{CodeChunk}
The result is given in Figure \ref{fig:NetW}.
The colorings convey that the core-structure consists of amine, oxidative stress and organic acid compounds mostly, while the periphery-structure consists of lipid compounds mostly.

\subsubsection{Analysis}
\begin{sloppypar}
Following visualization we might desire analyzing the retained structure.
The \fct{GGMnetworkStats} function calculates various network statistics, mostly at the node-level.
\end{sloppypar}
\begin{CodeChunk}
\begin{CodeInput}
R> NwkSTATS <- GGMnetworkStats(PcorP)
\end{CodeInput}
\end{CodeChunk}
It outputs an object of class \class{list} with, among others, slots for node degree (\code{$degree}) and node betweenness (\code{$betweenness}).
This information can, for example be used to evaluate which nodes sort high degree centrality scores.
\begin{CodeChunk}
\begin{CodeInput}
R> head(sort(NwkSTATS$degree, decreasing = TRUE))
\end{CodeInput}
\begin{CodeOutput}
    Amine_9    Amine_31    Amine_50  Org.Acid_5 Org.Acid_19 Org.Acid_21
          7           7           7           7           7           7
\end{CodeOutput}
\end{CodeChunk}
Or to evaluate which nodes sort high betweenness centrality scores.
\begin{CodeChunk}
\begin{CodeInput}
R> head(sort(NwkSTATS$betweenness, decreasing = TRUE))
\end{CodeInput}
\begin{CodeOutput}
Ox.Stress_13       Lip_64     Amine_50       Lip_50      Amine_2 Ox.Stress_29
    4943.551     4212.359     3021.298     2908.500     2789.298     2574.450
\end{CodeOutput}
\end{CodeChunk}
These results convey quantitatively what we grasped qualitatively from Figure \ref{fig:NetW}: Amines, and organic acids dominate the core structure (as reflected in the degree centralities).
Lipids \#50 and \#64 have high betweenness centralities as they act as gatekeepers between the core structure and the largest peripheral lipid structure.

\begin{figure}[t!]
\centering
\includegraphics[width=.9\textwidth]{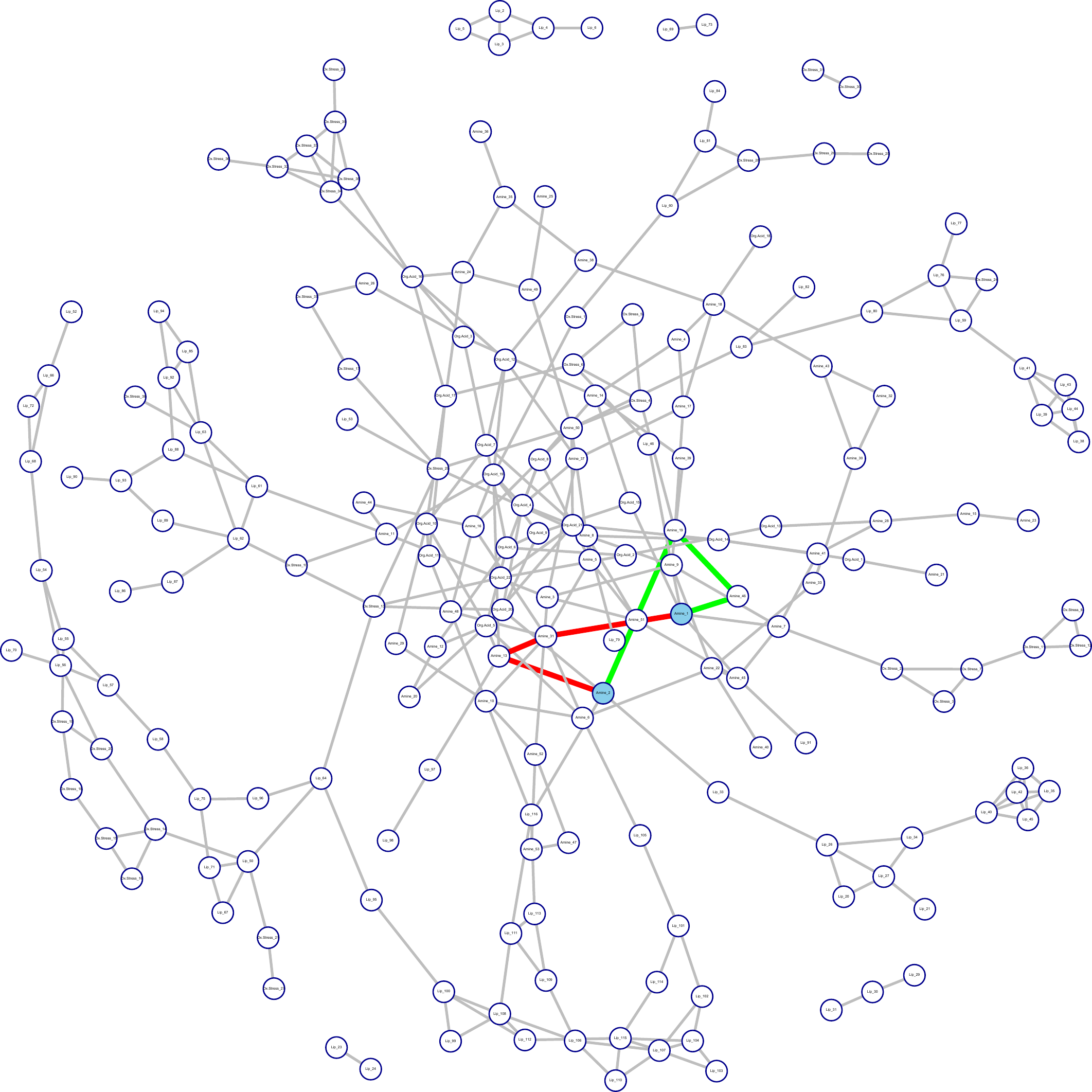}
\caption{\label{fig:Paths} The metabolite network of Figure \ref{fig:NetW}, but now visualized to convey the two strongest paths between Amines \#1 and \#2.
The nodes of interest are colored blue.
The edges of mediating paths are green while the edges of moderating paths are red.
The strongest mediating path is Amine \#1 $-$ Amine \#46 $-$ Amine \#19 $-$ Amine \#2.
The strongest moderating path is Amine \#1 $-$ Amine \#31 $-$ Amine \#13 $-$ Amine \#2.}
\end{figure}

We might also analyze network properties from the perspective of paths.
The \fct{GGMpathStats} function calculates path statistics for specified node pairs.
The arguments \code{node1} and \code{node2} are numerics designating the nodes of interest.
The function both calculates and visualizes the strongest paths.
The argument \code{nrPaths} then indicates how many paths should be visualized in the graphical output.
We could, for example, be interested in finding the two strongest paths between Amines \#1 and \#2.
\begin{CodeChunk}
\begin{CodeInput}
R> Paths <- GGMpathStats(P0$sparsePrecision,
+                        node1   = 1,
+                        node2   = 2,
+                        lay     = NULL,
+                        coords  = Coords,
+                        nrPaths = 2,
+                        Vsize   = 4,
+                        Vcex    = .2,
+                        prune   = TRUE,
+                        scale   = 2,
+                        legend  = FALSE)
\end{CodeInput}
\end{CodeChunk}
The graphical output, using the node-coordinates of previous calls to \fct{Ugraph}, can be found in Figure \ref{fig:Paths}.
As in \citet{JW05}, paths whose weights have an opposite sign to the marginal covariance (between endnodes of the path) are referred to as
\emph{moderating paths} while paths whose weights have the same sign as the marginal covariance are referred to as \emph{mediating} paths.
The edges of mediating paths are represented in green while the edges of moderating paths are represented in red.
Strong paths tend to be short paths.

\begin{figure}[h!]
\centering
\includegraphics[width=.8\textwidth]{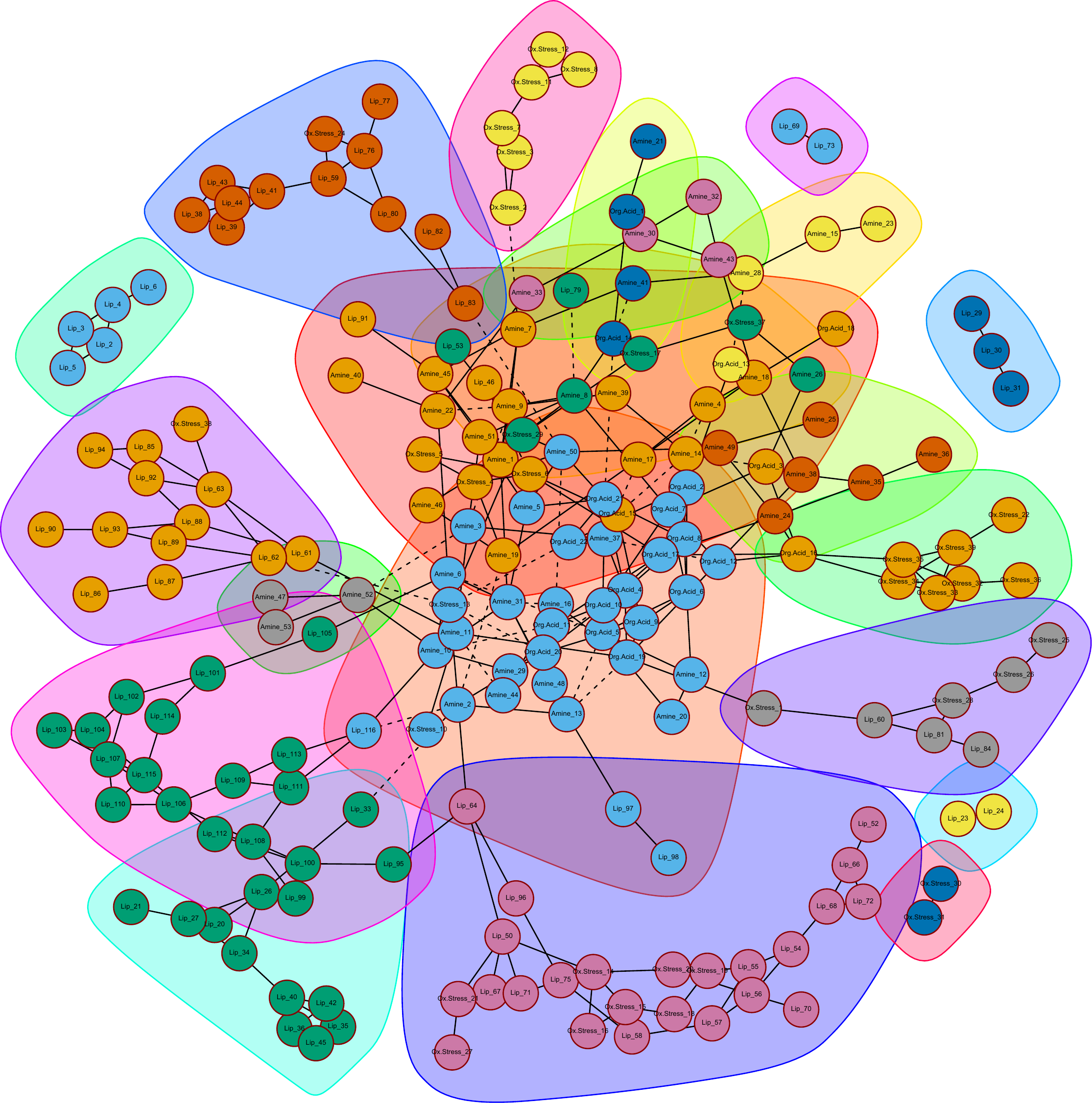}
\caption{\label{fig:Community} Community structure of the extracted metabolite network.}
\end{figure}

At the group-level we might be interested in detecting communities.
This can be done with the \fct{Communities} function.
When \code{graph = TRUE} (default) the community structure in the graph is also visualized.
The arguments for this visualization mirror a call to \fct{Ugraph}.
\begin{CodeChunk}
\begin{CodeInput}
R> set.seed(24354)
R> Commy <- Communities(PcorP, Vcolor = Colors, Vsize = 7,  Vcex = .3)
\end{CodeInput}
\end{CodeChunk}
The resulting visualized community structure can be found in Figure \ref{fig:Community}.
The same network now appears as was visualized in Figures \ref{fig:NetW} and \ref{fig:Paths}.
But now the network also expresses the community structure.
The colored borders demarcate communities.
The nodes are also colored according to community membership.
We see two large and intertwined amine-organic acid communities.
We also see various oxidative stress and lipid communities.
A numerical classification of module memberships for each node can be accessed through calling \code{Commy$membership}.

\subsection{Fused module} \label{SSec:FusedIllustrate}

\subsubsection{Extraction}
We could also consider the two classes of AD patients simultaneously.
We then take interest in jointly estimating the class-specific precision matrices as a basis for joint (or integrative) network modeling.
The data for AD class 1 patients (those without the genetic predisposition) consists of $n_1 = 40$ observations on $p = 230$ metabolic features.
First, we construct lists of class-specific target matrices and class-specific data matrices.
We scale the class-specific data, as the features have different scales and as the variability of the features may differ substantially.
Again, as we have little \emph{a priori} knowledge, we will keep the class-specific targets simple: Both classes are assigned an identity matrix as the target matrix.
\begin{CodeChunk}
\begin{CodeInput}
R> ADclass1 <- ADmetabolites[, sampleInfo$ApoEClass == "Class 1"]
R> ADclass2 <- ADmetabolites[, sampleInfo$ApoEClass == "Class 2"]
R> ADclass1 <- scale(t(ADclass1))
R> ADclass2 <- scale(t(ADclass2))
R> rAD1     <- cor(ADclass1)
R> rAD2     <- cor(ADclass2)
R> Rlist    <- list(rAD1 = rAD1, rAD2 = rAD2)
R> samps    <- c(nrow(ADclass1), nrow(ADclass2))
R> Tlist    <- default.target.fused(Slist = Rlist, ns = samps, type = "DUPV")
R> Ylist    <- list(AD1data = ADclass1, AD2data = ADclass2)
\end{CodeInput}
\end{CodeChunk}
The constructed lists will be directly used in the estimation of class-specific precision matrices.
As both classes consists of patients with AD, one would expect the network topologies of the respective groups to share some of their structure,
while potentially differing in a number of (topological) locations of interest.
The fusion framework of Section \ref{SSec:Fused} takes this into account explicitly.

The \fct{optPenalty.fused} function finds the optimal ridge and fusion penalty parameters via $K$-fold cross validation of the negative fused log-likelihood score.
It does so by using multi-dimensional optimization routines with automatically generated starting values.
The penalty-values at which this score is minimized are deemed optimal.
The call below uses $K = 10$ and a penalty structure specifying a ridge penalty for each class as well as a single fusion penalty (as we only have two classes in this simple example).
\begin{CodeChunk}
\begin{CodeInput}
R> set.seed(8910)
R> OPTf <- optPenalty.fused(
+    Ylist     = Ylist,
+    Tlist     = Tlist,
+    lambda    = as.matrix(cbind(c("ridge1", "fusion"),c("fusion", "ridge2"))),
+    cv.method = "kCV",
+    k         = 10,
+    verbose   = FALSE
)
\end{CodeInput}
\end{CodeChunk}
The class-specific precision matrices under the optimal penalties are contained in the list \code{OPTf$Plist}.
Again, it is informative to have a look at the penalties.
\begin{CodeChunk}
\begin{CodeInput}
R> OPTf$lambda.unique
\end{CodeInput}
\begin{CodeOutput}
      ridge1       fusion       ridge2
1.973739e+01 2.085230e-19 1.693029e+01
\end{CodeOutput}
\end{CodeChunk}
We see that the penalty values emphasize individual regularization over retainment of entry-wise similarities, indicating quite strong differences in class-specific precision matrices.
We could of course use the \fct{CNplot} function to see that both class-specific precision matrices are well-conditioned.

After obtaining class-specific precision matrices, we again need to determine their support in order to extract class-specific networks.
This is done through the \fct{sparsify.fused} function.
Its arguments are analogous to the arguments for the \fct{sparsify} function.
We again use lFDR thresholding and with \code{FDRcut} set to $.999$.
\begin{CodeChunk}
\begin{CodeInput}
R> P0s <- sparsify.fused(OPTf$Plist,
+                        threshold = "localFDR",
+                        FDRcut    = .999,
+                        verbose   = FALSE)
\end{CodeInput}
\begin{CodeOutput}
- Retained elements:  94
- Corresponding to 0.36 % of possible edges

- Retained elements:  289
- Corresponding to 1.1 % of possible edges
\end{CodeOutput}
\end{CodeChunk}
The \fct{sparsify.fused} function returns, for all classes considered, both the sparsified precision matrix and the sparsified partial correlation matrix.
For the Class 1 network, $94$ edges are retained, while for the Class 2 network again around $300$ edges are deemed to have a high probability of being present.
This could be a sign that the Class 1 network has a more erratic or random structure, making it harder to determine good estimates for the unknowns in the mixture distribution (\ref{eq:MIx}).
Visualization can support a first assessment.

\subsubsection{Visualization}
\begin{sidewaysfigure}
  \includegraphics[width=\textwidth]{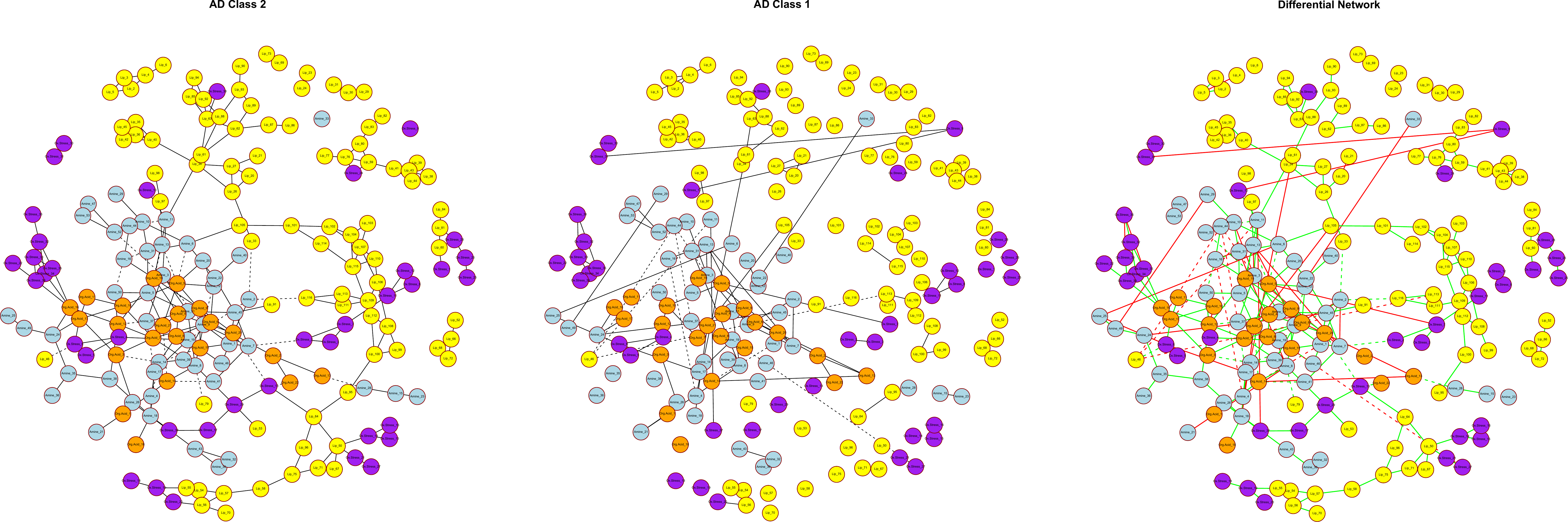}
    \caption{\emph{Left-hand panel}: Metabolite network for the AD Class 2 data.
    \emph{Middle-panel}: Metabolite network for the AD Class 1 data.
    \emph{Right-hand panel}: Differential network between AD classes 1 and 2.
    Dashed lines indicate negatively weighted edges while solid lines indicate positively weighted edges.
    Node colorings are according to compound family: blue for amines, orange for organic acids, yellow for lipids, and purple for oxidative stress compounds.
    Edges unique to Class 2 are depicted in green while edges unique to Class 1 are depicted in red.}
  \label{fig:NetWClass}
\end{sidewaysfigure}

When visualizing multiple networks over the same features, there are some additional issues to think of (in addition to layout, coloring, etc.).
One such issue is the possible retainment of coordinates over class-specific graphs.
The main function for visualization, \fct{Ugraph}, also supports visual comparison of class-specific networks.
Reusing the layout-coordinates of previous calls enables the visualization of class-specific networks in the same coordinates.
At times, this may be insightful as it allows one to visually track differential connections between class-specific networks.
In order to do so for the AD Class 1 and AD class 2 networks, we first use the \fct{Union} function.
This convenience function subsets two square matrices (over the same features) to the union of features
that have nonzero row (column) entries (i.e., to features implied in graphical connections).
This allows one to prune both networks to those features that are connected in at least one of them.
The AD Class 2 network will determine the coordinates for the AD Class 1 network.
Again, we use node-colorings to heighten the information-density of the visualization.
In addition, we use the \fct{DiffGraph} function, which visualizes in a single graph the edges that are unique to any two input networks.
It uses edge-colorings according to class-specific presence.
\begin{CodeChunk}
\begin{CodeInput}
R> TST <- Union(P0s$AD1data$sparseParCor, P0s$AD2data$sparseParCor)
R> PCclass1 <- TST$M1subset
R> PCclass2 <- TST$M2subset

R> Colors <- rownames(PCclass2)
R> Colors[grep("Amine", rownames(PCclass2))]     <- "lightblue"
R> Colors[grep("Org.Acid", rownames(PCclass2))]  <- "orange"
R> Colors[grep("Lip", rownames(PCclass2))]       <- "yellow"
R> Colors[grep("Ox.Stress", rownames(PCclass2))] <- "purple"

R> set.seed(111213)
R> par(mfrow=c(1,3))
R> Coords <- Ugraph(PCclass2,
+                   type   = "fancy",
+                   lay    = "layout_with_fr",
+                   Vcolor = Colors,
+                   prune  = FALSE,
+                   Vsize  = 7,
+                   Vcex   = .3,
+                   main   = "AD Class 2")
R> Ugraph(PCclass1,
+         type   = "fancy",
+         lay    = NULL,
+         coords = Coords,
+         Vcolor = Colors,
+         prune  = FALSE,
+         Vsize  = 7,
+         Vcex   = .3,
+         main   = "AD Class 1")
R> DiffGraph(PCclass1, PCclass2,
+            lay    = NULL,
+            coords = Coords,
+            Vcolor = Colors,
+            Vsize  = 7,
+            Vcex   = .3,
+            main   = "Differential Network")
\end{CodeInput}
\end{CodeChunk}
The resulting visualizations can be found in Figure \ref{fig:NetWClass}.
The left-hand panel gives the metabolite network for the Class 2 data.
It strongly resembles the network for the class 2 data obtained in Section \ref{SSec:CoreIllustrate}.
However, now we have a slightly lesser number of connections.
This is due to the fused estimation: There is some retainment of entry-wise similarities between the Class 1 and Class 2 networks.
The Class 1 network (middle-panel) seems more erratic (less structured) and less modular at first glance.
The differential network is given in the right-hand panel.
Edges unique to Class 2 are depicted in green while edges unique to Class 1 are depicted in red.
We can assess the seeming lack of structure of the Class 1 network vis-\`{a}-vis the Class 2 network with some simple analyzes.

\subsubsection{Analysis}
The analyzes at the node, path and community level performed in Section \ref{SSec:CoreIllustrate} have natural extensions to the situation in which we have multiple class-specific networks.
We will give examples for the node and community level.
In addition, a simple global (or network-level) analysis is performed.

The \fct{GGMnetworkStats.fused} function calculates various node-level statistics from a list of sparse precision/partial correlation matrices.
It returns a \class{data.frame} with the slotnames of the input list prefixed to the column-names.
This makes it relatively easy to, for example, compare the top node-degrees for class-specific networks.
\begin{CodeChunk}
\begin{CodeInput}
R> PC0list      <- list(PCclass1 = PCclass1, PCclass2 = PCclass2)
R> NwkSTATSList <- GGMnetworkStats.fused(PC0list)
R> DegreesAD1   <- data.frame(rownames(NwkSTATSList),
+                             NwkSTATSList$PCclass1.degree)
R> DegreesAD2   <- data.frame(rownames(NwkSTATSList),
+                             NwkSTATSList$PCclass2.degree)
R> DegreesAD1o  <- DegreesAD1[order(DegreesAD1[,2], decreasing = TRUE),]
R> DegreesAD2o  <- DegreesAD2[order(DegreesAD2[,2], decreasing = TRUE),]
R> head(DegreesAD1o, 7)
\end{CodeInput}
\begin{CodeOutput}
    rownames.NwkSTATSList. NwkSTATSList.PCclass1.degree
3                  Amine_3                            5
7                  Amine_7                            4
9                  Amine_9                            4
13                Amine_13                            4
74                   Lip_4                            4
167            Ox.Stress_6                            4
8                  Amine_8                            3
\end{CodeOutput}
\end{CodeChunk}
\begin{CodeChunk}
\begin{CodeInput}
R> head(DegreesAD2o, 7)
\end{CodeInput}
\begin{CodeOutput}
    rownames.NwkSTATSList. NwkSTATSList.PCclass2.degree
54              Org.Acid_5                            7
65             Org.Acid_16                            7
8                  Amine_8                            6
9                  Amine_9                            6
55              Org.Acid_6                            6
68             Org.Acid_19                            6
152                Lip_107                            6
\end{CodeOutput}
\end{CodeChunk}
At first glance the degree distributions seem to differ.
We will assess this with a simple test below.

\begin{figure}[h!]
\centering
\includegraphics[width=\textwidth]{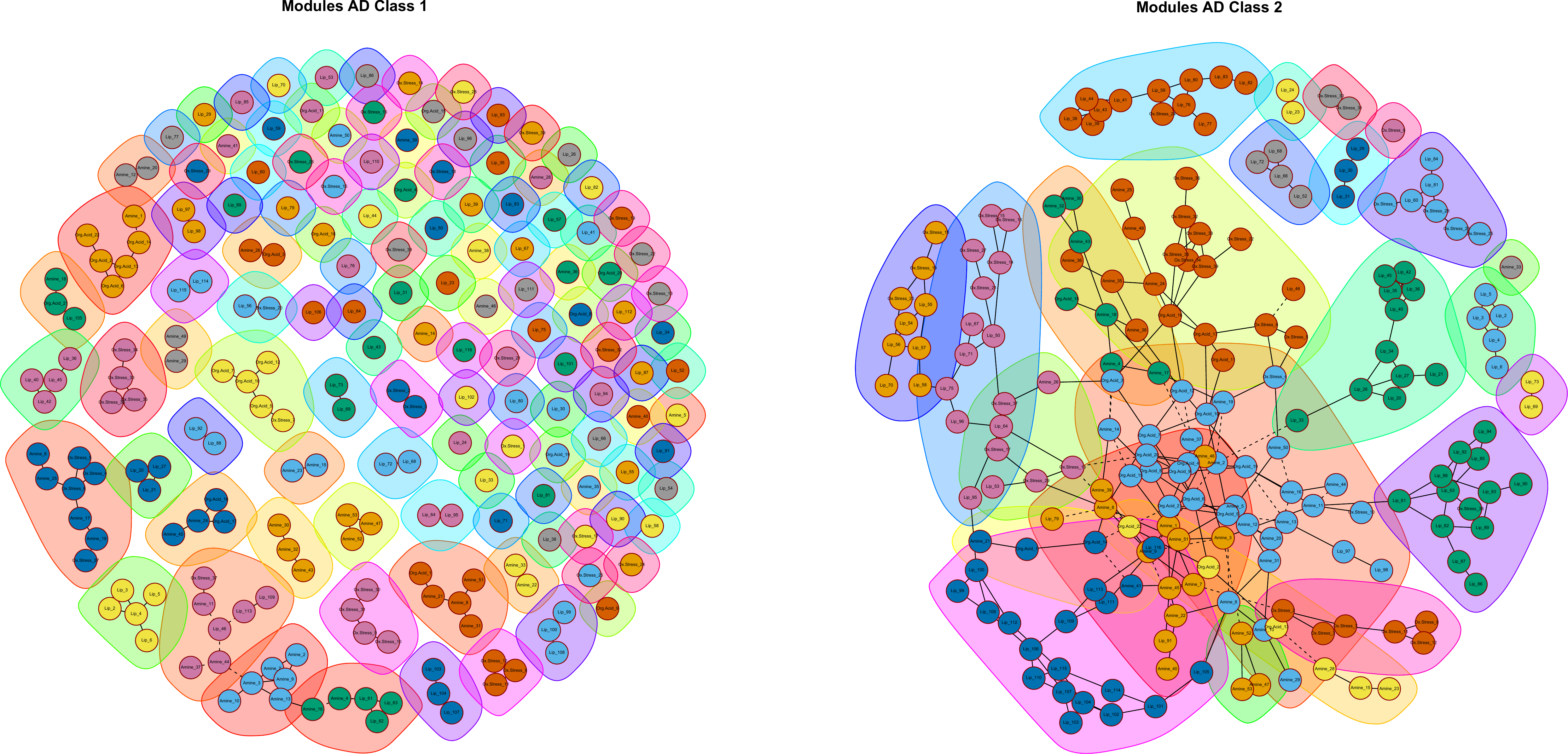}
\caption{\label{fig:CommunityClass} \emph{Left-hand panel}: Community structure for the AD Class 1 data.
\emph{right-hand panel}: Community structure for the AD Class 2 data.}
\end{figure}

We can apply the \fct{Communities} function to each class-specific sparsified matrix in order to find (and visualize) the communities for each class-specific network.
\begin{CodeChunk}
\begin{CodeInput}
R> set.seed(141516)
R> par(mfrow=c(1,2))
R> CommyC1 <- Communities(PCclass1, Vcolor = Colors, Vsize = 7,  Vcex = .3,
+                         main = "Modules AD Class 1")
R> CommyC2 <- Communities(PCclass2, Vcolor = Colors, Vsize = 7,  Vcex = .3,
+                         main = "Modules AD Class 2")
\end{CodeInput}
\end{CodeChunk}
The visual result is given in Figure \ref{fig:CommunityClass}.
The emerging picture is that the network for AD Class 2 is structured and modular while the network for AD Class 1 appears unstructured and loosely arranged.

Juxtaposing multiple networks also allows for tests on global network properties.
Simple testing could be done to assess differences in degree distributions or inherent randomness of the networks.
We can plot the degree distributions of the respective networks to see that these seem different.
\begin{CodeChunk}
\begin{CodeInput}
R> par(mfrow=c(1,1))
R> plot(density(DegreesAD1[,2]), col = "red", xlim = c(-1,8),
+       xlab = "Degree", main = "")
R> lines(density(DegreesAD2[,2]), col = "blue")
R> legenda <- c("AD class 1", "AD class 2")
R> legend(5, .5, legend = legenda, lwd = rep(1,2), lty = rep(1,2),
+         col = c("red", "blue"), cex = .7)
\end{CodeInput}
\end{CodeChunk}
\begin{figure}[h!]
\centering
\includegraphics[width=.83\textwidth]{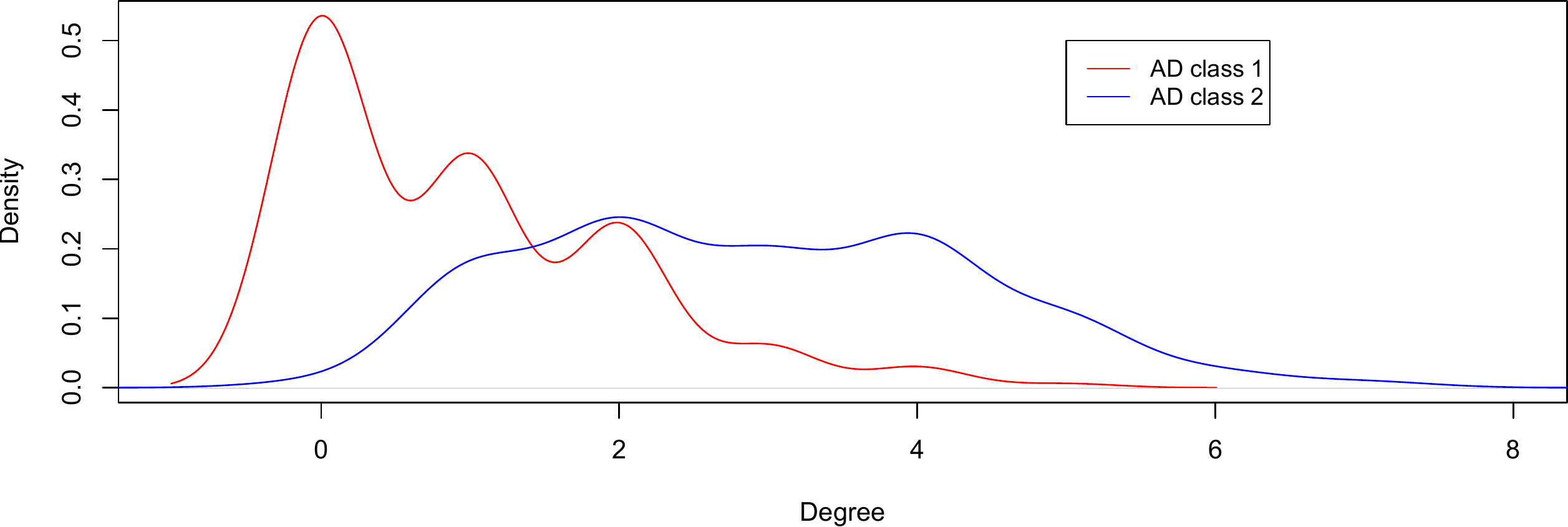}
\caption{\label{fig:Ddist} Degree distributions for the Class 1 and Class 2 networks.}
\end{figure}

For the Class 2 network, most nodes have a similar degree (Figure \ref{fig:Ddist}).
For the Class 1 network, the degree distribution seems to follow more of a power-law than a binomial distribution.
We can perform a simple dependent two-group Wilcoxon Signed Rank Test to see if the distribution of the difference in degrees is symmetric about 0.
We specify a one-sided alternative as we believe that the degree location of Class 2 is shifted to the right of the location for Class 1.
\begin{CodeChunk}
\begin{CodeInput}
R> wilcox.test(DegreesAD1[,2],DegreesAD2[,2], paired = TRUE,
+              alternative = "less")
\end{CodeInput}
\begin{CodeOutput}
	Wilcoxon signed rank test with continuity correction

data:  DegreesAD1[, 2] and DegreesAD2[, 2]
V = 444, p-value < 2.2e-16
alternative hypothesis: true location shift is less than 0
\end{CodeOutput}
\end{CodeChunk}
The results imply that this is indeed the case.
Hence, the degree distributions imply that the network for Class 1 is more loosely connected.
Class 2 has a known genetic predisposition for AD which may translate in relatively stable aberrations or amplifications across the metabolome.
Class 1 has no known genetic predisposition for AD.
The group is thus likely heterogeneous in disease aetiology, which may be a possible explanation for the lack of cohesiveness at the metabolic level.


\section{Summary and discussion}\label{Sec:Discuss}
The \pkg{rags2ridges} package was introduced in this paper.
It takes an $\ell_2$-approach towards graphical modeling of high-dimensional precision matrices.
This package supports the full network evaluation cycle, from extraction (through graphical modeling) to visualization and analysis.
It has a modular setup and technical details as well as usage of the core and fused modules were explicated above.
The core module revolves around the extraction, visualization, and analysis of single networks.
The fused module extends core methodology to multiple networks.
Another module is available: the chordal module.
This module extends core methodology to the extraction and analysis of graphs that can be triangulated (i.e., chordal graphs).
There is also a dependent sister package, \pkg{rag\underline{t}2ridges} \citep{ragt2ridges}, that uses the core module of \pkg{rags2ridges} for the graphical modeling of vector auto-regressive processes.

Future modular expansions will most likely consider functionality for robustification and directed graph inferral.
A robust module is conceivable in which the core technology is extended by relaxing the assumption of Gaussianity.
Another direction would be to conceive the sparsified precision or partial correlation matrix as a moralized representation of a directed graph.
One could then reverse-engineer the set of directed graphs that befit, in terms of conditional independence implications, the moralized representation.
We hope to offer such extensions in future updates of \pkg{rags2ridges}.
In addition, we are working towards making the package more \proglang{R}-idiomatic.


\section*{Computational details}
The results in this paper were obtained using \proglang{R}~4.0.2 with the \pkg{rags2ridges}~2.2.3 package.
\pkg{rags2ridges} imports from the following non-default packages:	\pkg{igraph}, \pkg{expm}, \pkg{reshape}, \pkg{ggplot2}, \pkg{Hmisc},
\pkg{fdrtool}, \pkg{snowfall}, \pkg{sfsmisc}, \pkg{gRbase}, \pkg{RBGL}, \pkg{graph}, \pkg{Rcpp}, and \pkg{RSpectra}.
\proglang{R} and all mentioned packages are available from the CRAN at \url{https://CRAN.R-project.org/}.

\section*{Acknowledgments}
This research was partly supported by Grant FP7-269553 (EpiRadBio) through the European Community's Seventh Framework Programme (FP7, 2007-2013).


\bibliography{refs}


\end{document}